\documentclass[twocolumn,showpacs,pre,aps]{revtex4}

\usepackage{dcolumn}

\begin{document}

\title{A Truth of Molecular Chaos}

\author{Yuriy E. Kuzovlev}
\email{kuzovlev@kinetic.ac.donetsk.ua}
\affiliation{A.A.Galkin Physics and Technology Institute
of NASU, 83114 Donetsk, Ukraine}


\begin{abstract}
The BBGKY hierarchy of equations for a particle interacting with an
ideal gas is investigated. Principal properties of its solutions are
disclosed, as exact identities which connect probability distribution
of path of the particle, its derivatives in respect to gas density
and irreducible many-particle correlations between gas molecules and
the path. They show that all the correlations always give equally
important contributions to evolution of the path distribution, and
therefore the exact theory does not reduce to the classical kinetics
even at arbitrary small gas density.
\end{abstract}


\maketitle

{\bf 1}. One of creators of the modern probability theory
A.\,Kolmogorov underscored \cite{kol} that in it ``\,{\it the concept
of independence of experiments fills most important place}\,'' and
``\,{\it correspondingly one of most important objectives of
philosophy of natural sciences}\,'' is ``\,{\it clearing-up and
refinement of those prerequisites under which one can treat given
phenomena as independent}\,''. Recall that in the probability theory
some random phenomena or quantities $\,A\,$ and $\,B\,$\, by
definition are independent if their probability distributions are
independent, that is $\,P(A,B)=P(A)\,P(B)\,$ \cite{kol}. However, in
natural sciences the independence of phenomena $\,A\,$ and $\,B\,$\,
is thought as absence of cause-and-effect relations between them,
that is an influence of one to another. Whether independence in this
usual sense does mean independence in the sense of the probability
theory?

From viewpoints of common sense and philosophy, certainly does not
mean. Merely because $\,A\,$ and $\,B\,$ which do not directly
influence one on another nevertheless both can be parts of a same
another random event and thus turn out to be indirectly related.

From the scientific point of view, it is natural to bring the same
question to the statistical mechanics. One of creators of modern
theory of dynamical systems and statistical mechanics N.\,Krylov
thoroughly analyzed it \cite{kr} and also came to the negative
answer: he concluded that opinions that ``\,{\it phenomena which are
``manifestly independent'' should have independent probability
distributions}\,'',\, and the like,\, are nothing but ``\,{\it
prejudices}\,'' \cite{kr}.

Especially Krylov pointed \cite{kr} to firmness of such prejudices
\cite{kr1}. Only it explains why the molecular chaos hypothesis put
forward by Boltzmann  many years ago \cite{bol} until now dominates
kinetics although never was logically substantiated \cite{dor}. And
why N.\,Bogolyubov, when he obtained \cite{bog} an exact hierarchy of
evolution equations for $\,s\,$-particle distribution functions,
straight away truncated his equations at $\,s=2\,$ thus reducing it
to the Boltzmann equation.

Undoubtedly, molecules of sufficiently rarefied gas are independent
in usual sense since almost surely have nothing common in the past.
Nevertheless they can be essentially dependent in the sense of the
probability theory. This is quite understandable \cite{i1} (or you
may see \cite{i2}). The absence of common causes of colliding
particles in the past means, for each of them, absence of any back
reaction of the gas to its past collisions. Therefore arbitrary long
fluctuations in relative frequency of collisions are allowable
\cite{kr2}. These fluctuations just play the role of aforesaid random
events producing indirect statistical interdependencies between pairs
(or groups) of molecules capable of being participators of one and
the same collision (or a cluster of successive collisions).

As the consequence, $\,P(A,B)  \neq P(A)\,P(B)\,$ where $\,P(A)\,$ is
probability of finding a molecule at (phase) point $\,A\,$ and
$\,P(A,B)\,$ is probability of finding simultaneously two molecules
at points $\,A\,$ and $\,B\,$. At that, relaxation of one-particle
distribution $\,P(A)\,$ is determined by pair correlation $\,P(A,B) -
P(A)\,P(B)\,$. Relaxation of the latter just similarly always
(regardless of the gas rarefaction) is determined by three-particle
correlation. And so on up to infinity. Since during time interval
$\,t\,$ a molecular undergoes $\,\sim t/\tau \,$ collisions (with
$\,\tau\,$ being characteristic free-flight time), a correct
description of gas evolution over this interval requires taking into
account $\,s$-particle correlations with at least\, $\,s \lesssim
t/\tau \,$. Hence, in practice the whole hierarchy of equations
deduced by Bogolyubov \cite{bog} is necessary.

In work \cite{i1} (and additionally or instead in \cite{i2}) and in
work \cite{p1} approximate solutions to this hierarchy or, in other
words, the Bogolyubov-Born-Green-Kirkwood-Yvon (BBGKY) equations
\cite{re} were suggested for the problem about random wandering of a
test molecule, and explanations were expounded why the Boltzmann's
hypothesis is wrong. The aim of the present communication is to prove
the statements of preceding paragraph without any approximations. At
that we will strengthen the proof and besides simplify it due to
replacing the usual gas by ideal gas whose molecules interact with
the test molecule only but not with each other.

\,\,\,

{\bf 2}. We want to consider thermal random motion of a test molecule
(TM) in thermodynamically equilibrium gas, at that specifying its
position $\,{\bf R}(t)\,$ at some initial time moment $\,t=0\,$:\,
$\,{\bf R}(0)={\bf R}_0\,$.

Let $\,{\bf P}\,$ and $\,M\,$ denote momentum and mass of TM,
$\,m\,$, $\,{\bf r}_j\,$ and $\,{\bf p}_j\,$ ($\,j=1,2,...\,$) denote
masses, coordinates and momenta of other molecules,\, $\,\Phi({\bf
r})\,$ is (short-range repulsive) potential of interaction between
any of them and TM, and $\,n\,$ is gas density (mean concentration of
molecules). At arbitrary time $\,t\geq 0\,$, full statistical
description of this system is presented by the chain of
$\,(k+1)$-particle distribution functions ($\,k=0,1,2,...\,$):\,
$\,F_0(t,{\bf R},{\bf P}|\,{\bf R}_0\,;n\,)\,$\, which is normalized
(to unit) density of probability distribution of TM's variables, and
\,$\,F_k(t,{\bf R}, {\bf r}^{(k)},{\bf P},{\bf p}^{(k)}|\,{\bf
R}_0\,;n\,)\,$\, (where\, $\,{\bf r}^{(k)} =\{{\bf r}_1...\,{\bf
r}_k\,\}\,$, $\,{\bf p}^{(k)} =\{{\bf p}_1...\,{\bf p}_k\,\}\,$)
which are probability densities of finding TM at point $\,{\bf R}\,$
with momentum $\,{\bf P}\,$ and simultaneously finding out some
$\,k\,$ molecules at points $\,{\bf r}_j\,$ with momenta $\,{\bf
p}_j\,$. A rigorous definition of such distribution functions (DF)
was done in \cite{bog}. In respect to the coordinates $\,{\bf r}_j\,$
they are not normalized, but instead (as in \cite{bog}) obey the
conditions of weakening of inter-particle correlations under spatial
separation of particles. Subject to the symmetry of DF in respect to
$\,x_j= \{{\bf r}_j,{\bf p}_j\}\,$\, these conditions can be
compactly written as follows:\,
$\,F_k\,\rightarrow\,F_{k-1}\,G_m({\bf p}_k)\,$\, at \,$\,{\bf
r}_k\rightarrow\infty\,$\,,\, where $\,G_m({\bf p})\,$ is the Maxwell
momentum distribution of a particle with mass $\,m\,$.

The enumerated DF satisfy a standard chain of the Bogolyubov
equations \cite{bog}:
\begin{equation}
\frac {\partial F_k}{\partial t}=[\,H_{k},F_k\,]+n \,\frac {\partial
}{\partial {\bf P}}\int_{k+1}\!\!\Phi^{\,\prime}({\bf R}-{\bf
r}_{k+1}) \,F_{k+1}\,\label{fn}
\end{equation}
($\,k=0,1, ...\,$) along with obvious initial conditions
\begin{equation}
\begin{array}{c}
F_k|_{t=\,0}\,=\delta({\bf R}-{\bf R}_0)\, \exp{(-H_k/T\,)}= \label{ic}\\
= \delta({\bf R}-{\bf R}_0)\,G_M({\bf P}) \prod_{j\,=1}^k E({\bf
r}_j-{\bf R})\, G_m({\bf p}_j)\,\,,
\end{array}
\end{equation}
where\, $\,H_{k}\,$ is Hamiltonian of subsystem ``\,$k$ molecules +
TM\,'',\, $\,\int_k ... =\int\int ...\,\,d{\bf r}_k\,d{\bf
p}_k\,$\,,\, $\,[...,...]\,$ means the Poisson brackets,\,
$\,\Phi^{\,\prime}({\bf r}) =\nabla\Phi({\bf r})\,$\,, and $\,E({\bf
r})=\exp{[-\Phi({\bf r})/T\,]}\,$. Notice that TM can be considered
as a molecule of non-uniformly distributed impurity, and equations
(\ref{fn}) are identical to the equations of two-component gas
\cite{bog} in the limit of infinitely rare impurity, when the main
component is spatially homogeneous and thermodynamically equilibrium.

Equations (\ref{fn}) together with (\ref{ic}) unambiguously determine
evolution of $\,F_0\,$ and eventually probability distribution of
TM's displacement $\,{\bf R}-{\bf R}_0\,$. These equations will
become more clear if we make a linear change of DF $\,F_k\,$ by new
functions $\,V_k\,$ with the help of recurrent relations as follow:
\[
\begin{array}{c}
F_0(t,{\bf R},{\bf P}| \,{\bf R}_0;n)\,=\,V_0(t,{\bf R},{\bf P}|\,
{\bf R}_0;n)\,\,\,,
\end{array}
\]
\begin{equation}
\begin{array}{c}
F_1(t,{\bf R},{\bf r}_1,{\bf P},{\bf p}_1|\,{\bf
R}_0;n)\,=\,\\
=\,V_0(t,{\bf R},{\bf P}|\,{\bf R}_0;n)\,f({\bf r}_1\!-{\bf R},{\bf
p}_1)\,+\\+\, V_1(t,{\bf R},{\bf r}_1,{\bf P},{\bf p}_1|\,{\bf
R}_0;n)\,\,\,, \label{cf1}
\end{array}
\end{equation}
where\, $\,f({\bf r},{\bf p}) = E({\bf r})\,G_m({\bf p})\,$\,,

\[
\begin{array}{c}
F_2(t,{\bf R},{\bf r}^{(2)},{\bf P},{\bf p}^{(2)}|{\bf
R}_0;n)\,= \,\\
=V_0(t,{\bf R},{\bf P}|{\bf R}_0;n) f({\bf r}_1\!-{\bf R},{\bf
p}_1)f({\bf r}_2\!-{\bf R},{\bf p}_2)\,+\\
+\, V_1(t,{\bf R},{\bf r}_1,{\bf P},{\bf p}_1|\,{\bf R}_0;n)\, f({\bf
r}_2\!-{\bf R},{\bf p}_2)\,+\\
+\, V_1(t,{\bf R},{\bf r}_2,{\bf P},{\bf p}_2|\,{\bf R}_0;n)\, f({\bf
r}_1\!-{\bf R},{\bf p}_1)\,+\\
+ \,V_2(t,{\bf R},{\bf r}^{(2)}, {\bf P},{\bf p}^{(2)}| \,{\bf
R}_0;n)\,\,\,,
\end{array}
\]
and so on.

Apparently, from viewpoint of the probability theory,\, $\, V_k\,$
represent a kind of cumulants (semi-invariants), or cumulant
functions (CF). It is important to notice that zero values of these
CF would mean that all conditional DF of gas, $\,F_k/F_0\,$,\, are
independent on initial position $\,{\bf R}_0\,$ of TM and thus on its
displacement $\,{\bf R}-{\bf R}_0\,$. This fact makes visible an
interesting specificity of CF\, $\,V_k\,$\,:\, they are irreducible
correlations of not only current dynamic states of TM and $\,k\,$ gas
molecules but also correlations of all them with total previous TM's
displacement.

In terms of the CF the BBGKY hierarchy acquires a more complicated
tridiagonal structure (we omit uninteresting algebraic details):
\begin{eqnarray}
\frac {\partial V_{k}}{\partial t}=[H_k,V_k]+n \,\frac {\partial
}{\partial {\bf P}}\int_{k+1}\!\! \Phi^{\prime}({\bf
R}-{\bf r}_{k+1})V_{k+1}+\nonumber\\
+\,T\sum_{j\,=1}^{k}\, \mathrm{P}_{kj}\,G_m({\bf p}_k)
\,E^{\prime}({\bf r}_k-{\bf R}) \left[\frac {{\bf P}}{MT}+\frac
{\partial }{\partial {\bf P}}\right ] V_{k-1}\,\,\label{vn}
\end{eqnarray}
Here\, $\,E^{\prime}({\bf r})=\nabla E({\bf r})\,$,\, and\,
$\,\mathrm{P}_{kj}\,$\, symbolizes transposition of the pairs of
arguments $\,x_j\,$\, and $\,x_k\,$. On the other hand, initial
conditions (\ref{ic}) and the above-mentioned conditions of weakening
of correlations \cite{bog} take very simple form:
\begin{equation}
\begin{array}{c}
V_0(0\,,{\bf R},{\bf P}|\,{\bf R}_0;\,n)\,=\delta({\bf R}-{\bf
R}_0) \,G_M({\bf P})\,\,,\\
V_{k}(0\,,{\bf R}, {\bf r}^{(k)},{\bf P},{\bf p}^{(k)}|\,{\bf
R}_0;n)=0\,\,,\label{icv}\\
V_k(t,{\bf R}, {\bf r}^{(k)},{\bf P},{\bf p}^{(k)}|{\bf
R}_0;n)\rightarrow0\,\,\,\,\,\,\texttt{at}\,\,\,\,{\bf
r}_j\rightarrow \infty
\end{array}
\end{equation}
($\,1\leq j\leq k\,$). Thus, as it should be with cumulants, CF
$\,V_k\,$ disappear under removal of even one of molecules.

From these equations and initial conditions (as well as physical
reasonings) it is clear that the reduction to zero in (\ref{icv})
realizes in an integrable way, so that integrals\,
$\,\overline{V}_{k}=\int_{k+1} V_{k+1}\,$\, are finite. Let us
consider them. By applying the operation $\,\int_k\,$ to equations
(\ref{vn}) one easy obtains equations
\begin{eqnarray}
\frac {\partial \overline{V}_{k}}{\partial t}=[H_k,\overline{V}_k]+n
\,\frac {\partial }{\partial {\bf P}}\int_{k+1}\!\!
\Phi^{\prime}({\bf
R}-{\bf r}_{k+1})\,\overline{V}_{k+1}+\nonumber\\
+\,\frac {\partial }{\partial {\bf P}}\int_{k+1}\!\!
\Phi^{\prime}({\bf
R}-{\bf r}_{k+1})\,V_{k+1}\,+\,\,\,\,\,\,\,\,\, \,\, \label{vn1}\\
+\,T\sum_{j\,=1}^{k}\, \mathrm{P}_{kj}\,G_m({\bf p}_k)
\,E^{\prime}({\bf r}_k-{\bf R}) \left[\frac {{\bf P}}{MT}+\frac
{\partial }{\partial {\bf P}}\right ] \overline{V}_{k-1}\nonumber
\end{eqnarray}
(with\, $\,k=0,1,...\,$). Because of (\ref{icv}) initial conditions
to these equations are zero:\, $\,\overline{V}_{k}(t=0)=0\,$\, at any
$\,k\,$.

Now, in addition to $\,\overline{V}_{k}\,$, let us consider
derivatives of CF in respect to the gas density,\, $\,V^{\prime}_{k}=
\partial V_{k}/\partial n\,$\,. It is easy to see that differentiation
of (\ref{vn}) in respect to $\,n\,$ yields equations for the
$\,V^{\prime}_{k}\,$ which exactly coincide with (\ref{vn1}) after
changing there $\,\overline{V}_{k}\,$ by $\,V^{\prime}_{k}\,$.
Besides, in view of (\ref{icv}), initial conditions to these
equations again all are zero:\, $\,V^{\prime}_{k}(t=0)=0\,$\, at any
$\,k\geq 0\,$. These observations strictly imply exact equalities\,
$\,V^{\prime}_{k}=\overline{V}_{k}\,$,\, or
\begin{eqnarray}
\frac {\partial }{\partial n}\,\, V_{k}(t,{\bf R}, {\bf r}^{(k)},{\bf
P},{\bf p}^{(k)}|\,{\bf R}_0;n)\,=\,\, \,\,\,\, \label{me}\\
=\,\int_{k+1} V_{k+1}(t,{\bf R}, {\bf r}^{(k+1)},{\bf P},{\bf
p}^{(k+1)}|\,{\bf R}_0;n)\,\,\nonumber
\end{eqnarray}
This is main formal result of the present paper.

\,\,\,

{\bf 3}. The result (\ref{me}) contains the proof promised in Sec.1.
Indeed, equalities (\ref{me}) show, firstly, that all the
many-particle correlations between gas molecules and past
displacement of test molecule (TM) really exist, i.e. differ from
zero. Secondly, all they have roughly one and the same order of
magnitude. For instance, if comparing their integral values, due to
(\ref{me}) we can write, in natural dimensionless units,
\[
n^k\!\int_1\!\! ...\!\int_k \int\! V_{k}\,d{\bf P}\,=\,
n^kV_0^{(k)}(t,\Delta;n) \,\sim\, c_k V_0(t,\Delta;n)\,\,,
\]
where\, $\,V_0(t,\Delta;n)  =\int V_0(t,{\bf R},{\bf P}|{\bf
R}_0;n)\,d{\bf P}\,$\, is probability distribution of the TM's
displacement\, $\,\Delta ={\bf R}-{\bf R}_0\,$\,,
\,$\,V_0^{(k)}(t,\Delta;n)=\partial ^{\,k}V_0(t,\Delta;n)/\partial
n^k\,$\, are its derivatives in respect to gas density $\,n\,$\,,
and\, $\,c_k\,$ some numeric coefficients. Hence, all the
correlations are equally important, and none of them can be neglected
if we aim at knowledge about true statistics of TM's random walk.

For more details, let us suppose that $\,(s+1)$-particle correlation
is so insignificant that one can assign $\,V_s=0\,$ in (\ref{vn}). At
that, according to (\ref{vn})-(\ref{icv}), all higher-order
correlations also will be rejected. Then, obviously, according to
(\ref{me}), distribution $\,V_0(t,{\bf R},{\bf P}|{\bf R}_0;n)\,$ and
thus $\,V_0(t,\Delta;n)\,$ must depend on\, $\,n\,$\, definitely as
an $\,(s-1)$-order polynomial. But, from the other hand,
distribution\, $\,V_0\,$\, what follows from the truncated chain of
equations (\ref{vn}) certainly is absolutely non-polynomial function
of\, $\,n\,$\,. With taking into account that equalities (\ref{me})
do express exact properties of solutions to (\ref{vn})-(\ref{icv}) we
see that very deep contradiction is on hand.

This contradiction clearly prompts us that truncation of the BBGKY
hierarchy leads to qualitative losses in its solution.

Some possible losses already were characterized in \cite{i1} and
\cite{i2} (and firstly even much earlier in \cite{bk12}) and in part
filled up in \cite{i1,p1}. Therefore here we confine ourselves
(continuing 5-th paragraph of Sec.1) by remark that cutting of the
$\,(s+1)$-particle correlation means cutting of $\,s$-th and higher
statistical moments of fluctuations in relative frequency of TM's
collisions with gas molecules (in other words, fluctuations in
diffusivity of TM \cite{bk12}). At\, $\,s=2\,$ these fluctuations are
completely ignored, and such truncated equations (\ref{vn}) yield a
closed equation for $\,V_0(t,{\bf R},{\bf P}|{\bf R}_0;n)\,$ which is
equivalent to the Boltzmann-Lorentz equation \cite{re}.

It is necessary to emphasize that above reasonings, as well as the
exact relations (\ref{me}), are indifferent to a degree of the gas
rarefaction. Consequently, one can state that the Boltzmann-Lorentz
equation (moreover, all the classical kinetics including the
Boltzmann equation and its generalizations) does not represent a
(low-density) limit of the exact statistical mechanical theory. The
conventional kinetics is only (more or less adequate or caricature)
probabilistic model of exact theory. Of course, in the latter also
molecular chaos does prevail. But here it is much more rich, even if
speaking about rarefied gas, and does not keep within naive
probabilistic logics.


\begin{thebibliography}{12}
\bibitem{kol}
A.\,N.\,Kolmogorov.\, Foundations of the theory of probability.
Chelsea, New York, 1956.

\bibitem{kr}
N.\,S.\,\,Krylov.\, Works on the foundations of statistical
physics.\, Princeton, 1979.

\bibitem{kr1}
These prejudices ``{\it are so habitual that even persons who agreed
with our argumentation usually automatically go back to them when
facing with a new question\,}'' \cite{kr}.

\bibitem{bol}
L.\,Boltzmann.\,  Vorlesungen uber Gastheorie.\, Bd. 1-2.\, Leipzig,
1896-1898.

\bibitem{dor}
J.\,R.\,Dorfman.\, An introduction to chaos in non-equilibrium
statistical mechanics.\, Cambridge, 1999.

\bibitem{bog}
N.\,N.\,\,Bogolyubov.\, Problems of dynamical theory in statistical
physics.\, North-Holland, 1962.

\bibitem{i1}
Yu.\,E. Kuzovlev,\, Sov.Phys.\,-\,JETP\, {\bf 67},\, \,No.\,12,\,
2469 (1988).

\bibitem{i2}
Yu.\,E.\,Kuzovlev, arXiv:\, cond-mat/9903350\,.

\bibitem{kr2}
``\,{\it ... relative frequencies of some phenomenon along a given
phase trajectory, generally speaking, in no way are connected to
probabilities}\,'' \cite{kr}.

\bibitem{p1}
Yu.\,E.\,Kuzovlev, arXiv:\, cond-mat/0609515, 0612325\,.

\bibitem{re}
P.\,Resibois and M.\,de\,Leener. Classical kinetic theory of fluids.
Wiley, New-York, 1977.

\bibitem{bk12}
Yu.\,E. Kuzovlev and G.\,N. Bochkov,\, Radiophysics and Quantum
Electronics\, {\bf 26},\, No.\,3,\, 228 (1983);\, {\bf 27},\, No.9
(1984).

\end{thebibliography}
\end{document}